# A sluggish mid-Proterozoic biosphere and its effect on Earth's redox balance

Kazumi Ozaki[1,2,3]*, Christopher T. Reinhard[1,2], and Eiichi Tajika[4]

[1]School of Earth and Atmospheric Sciences, Georgia Institute of Technology, Atlanta, GA 30332, USA

[2]NASA Astrobiology Institute, Alternative Earths Team. Mountain View, CA 94043, USA

[3]NASA Postdoctoral Program, Universities Space Research Association, Columbia, MD 21046, USA

[4]Department of Earth and Planetary Science, Graduate School of Science, The University of Tokyo, Bunkyo-ku, Tokyo 113-0033, Japan

*Corresponding author: E-mail: kazumi.ozaki@eas.gatech.edu

**Abstract:** The possibility of low but non-trivial atmospheric oxygen ($O_2$) levels during the mid-Proterozoic (between 1.8 and 0.8 billion years ago, Ga) has important ramifications for understanding Earth's $O_2$ cycle, the evolution of complex life, and evolving climate stability. However, the regulatory mechanisms and redox fluxes required to stabilize these $O_2$ levels in the face of continued biological oxygen production remain uncertain. Here, we develop a biogeochemical model of the C-N-P-$O_2$-S cycles and use it to constrain global redox balance in the mid-Proterozoic ocean-atmosphere system. By employing a Monte Carlo approach bounded by observations from the geologic record, we infer that the rate of net biospheric $O_2$ production was $3.5_{-1.1}^{+1.4}$ Tmol yr$^{-1}$ (1$\sigma$), or ~25% of today's value, owing largely to phosphorus scarcity in the ocean interior. Pyrite burial in marine sediments would have represented a comparable or more significant $O_2$ source than organic carbon burial, implying a potentially important role for Earth's sulphur cycle in balancing the oxygen cycle and regulating atmospheric $O_2$ levels. Our statistical approach provides a uniquely comprehensive view of Earth system biogeochemistry and global $O_2$ cycling during mid-Proterozoic time and implicates severe P biolimitation as the backdrop for Precambrian geochemical and biological evolution.

**Summary:** Biogeochemical model inversion provides persuasive evidence for a strongly suppressed ('sluggish') oxygenic biosphere as the primary cause of Earth's protracted oxygenation.

# 1 INTRODUCTION

The modern Earth's biosphere generates copious amounts of molecular oxygen ($O_2$)—the results of which can be seen from the uppermost atmosphere to the deepest reaches of the ocean. This remarkable abundance of $O_2$ is ultimately sustained by the activity of oxygenic photosynthesis, representing a critical control on the evolution of surface environments throughout the Earth's history and the emergence and expansion of biological complexity. Although this history has been reconstructed in broad strokes, significant gaps remain in our understanding of the cause-and-effect relationships and quantitative features regulating Earth system evolution. For example, geochemical proxies are consistent with a broad range of atmospheric $O_2$ levels during Earth's middle age (the 'mid-Proterozoic', between ~1.8 to 0.8 billion years ago, Ga) (Holland, 1984; Kump, 2008; Lyons et al., 2014) and are themselves agnostic as to the mechanisms regulating the global $O_2$ cycle. From this point of view, unravelling the global $O_2$ budget during mid-Proterozoic time is crucial as a window into planetary $O_2$ cycling, early eukaryotic evolution and the ultimate rise of animals.

A range of sedimentological and geochemical observations (Cole et al., 2016; Hardisty et al., 2017; Planavsky et al., 2014; Tang et al., 2016) have suggested the intriguing possibility that atmospheric $O_2$ levels remained low (<0.1–10% of the present atmospheric level, PAL) for much of mid-Proterozoic time, yet the mechanism(s) and redox fluxes involved in maintaining such low $O_2$ levels remain enigmatic. In particular, the fragmentary nature of Earth's Precambrian rock record and uncertainty in isotope mass balance models (Bjerrum and Canfield, 2004; Hayes and Waldbauer, 2006; Krissansen-Totton et al., 2015; Schrag et al., 2013) has resulted in concomitant uncertainties in Earth's evolving $O_2$ budget. One simple explanation would be a less active oxygenic biosphere, perhaps mediated through limitation of oxygenic photosynthesis by major nutrients (P and N) (Bjerrum and Canfield, 2002; Derry, 2015; Laakso and Schrag, 2014; Planavsky et al., 2010; Reinhard et al., 2017) and bioessential trace elements (Anbar and Knoll, 2002; Scott et al., 2008). However, fully elucidating the factors linking nutrient availability, the size and scope of Earth's biosphere, and atmospheric $O_2$ during the mid-Proterozoic requires an explicitly quantitative biogeochemical framework for $O_2$ production and consumption that satisfies global redox balance and is consistent with constraints from the geologic record.

## 2 METHODS

Here, we employ a novel biogeochemical model of the ocean-atmosphere system (CANOPS) with the goal of statistically constraining the rate of biotic $O_2$ production and the attendant large-scale biogeochemistry of the mid-Proterozoic Earth system. CANOPS is a 1-D (vertically resolved) intermediate complexity box model of ocean biogeochemistry (Ozaki and Tajika, 2013; Ozaki et al., 2011; Reinhard et al., 2017) (see Fig. 1 and Supplementary Methods for a further explanation). We use constraints from the geologic record to perform a Monte Carlo analysis of significant biogeochemical fluxes based on a wide range of values for poorly constrained control parameters (see Table 1 and Supplementary Methods), which offers a comprehensive and robust quantitative picture of Earth system biogeochemistry. In this study we focus on eight key parameters that play a quantitatively important role in the marine carbon, phosphorus and sulphur budgets. For the sake of objective analysis, we adopt uniform (non-informative) prior distributions, except for crustal sulphur reservoirs, for which Gaussian probability density functions were assumed. Each simulation was run long enough to establish a steady state, verified by looking at the evolution of the concentration of chemical species and the sulphur budget over tens of millions of years.

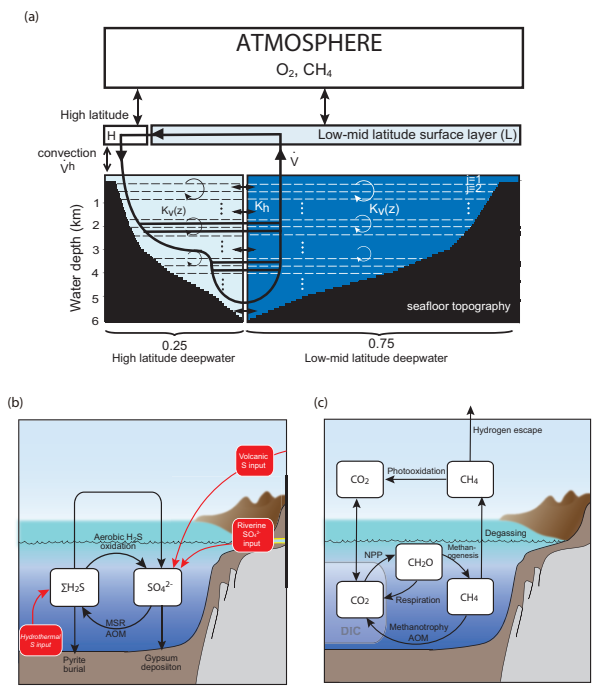

**FIGURE 1** Model schematic. (a) CANOPS model configuration. The parameters regarding geometry and water transport are tabulated in Table S3. (b) Global sulphur cycle schematic. Two sulphur species ($SO_4^{2-}$ and $\Sigma H_2S$) are transformed each other via microbial sulphate reduction (MSR), AOM, and sulphide oxidation reactions. Riverine and subaerial/submarine volcanic inputs are the primary source of sulphur to the ocean, and the burial of pyrite and gypsum in marine sediments is the primary sink. It is assumed that hydrogen sulphide escaping from the ocean to the atmosphere is completely oxidized and returns to the ocean as sulphate. The organic sulphur cycle is ignored in this study. (c) Global $CH_4$ cycle schematic. The model includes $CH_4$ generation via methanogenesis and its oxidation reactions via methanotrophy and AOM in the ocean interior, as well as $CH_4$ degassing flux to the atmosphere and its photooxidation. The rates of $CH_4$ photooxidation and hydrogen escape to space are calculated according to the parameterizations proposed by (Goldblatt et al., 2006). No input of abiotic $CH_4$ via continental hydrothermal systems is accounted for because previous estimate of modern value (<0.3 Tmol yr$^{-1}$; (Fiebig et al., 2009)) is negligible relative to the biological flux, although we realize that it would play a role in the global redox budget.

We sample and statistically analyse a subset of model runs that yield seawater sulphate ($SO_4^{2-}$) concentrations between 0.1–1 mM, as constrained by the stable isotope record of Proterozoic marine sedimentary rocks (Luo et al., 2015; Lyons and Gill, 2010; Planavsky et al., 2012; Scott et al., 2014). Of the total ~22,000 simulations performed, 3,342 simulations met this criterion (the 'All' scenario of Table S5). In our 'Low $O_2$' and 'High $O_2$' scenarios, we assumed lower upper limits for $f_{erosion}$ and $K_{MSR}$, yielding 621 and 308 simulations for analysis. We tracked the median value of $J_{oc}^{bur}$ and $J_{py}^{bur}$ to evaluate convergence, e.g. whether the sampling procedure was run for long enough to obtain the stationary distribution (Fig. S1). Like all geochemical constraints, mid-Proterozoic $SO_4^{2-}$ levels are subject to some degree of uncertainty. However, the magnitude of uncertainty is significantly less than that of other poorly constrained boundary conditions, and in any case our principal conclusions are not strongly affected by relaxing this constraint. For example, the upper bound could be higher (Kah et al., 2004), but in this case the model retrieves parameter combinations that predict lower biological productivity, strengthening our core arguments. This arises as a result of the need to supress microbial sulphate reduction (MSR) (and subsequent pyrite precipitation) in order to achieve higher [$SO_4^{2-}$], which is most straightforwardly achieved by a decrease in the availability of organic matter for MSR.

## 3 RESULTS AND DISCUSSION

### 3.1 Limited $O_2$ production in Proterozoic oceans

The most striking result of our statistical analysis is the need for a limited rate of $O_2$ production from organic carbon and pyrite sulphur burial in order to remain consistent with existing constraints on seawater [$SO_4^{2-}$]. For our 'Low $O_2$' scenario, the posterior probability distribution for the global rate of organic carbon burial ($J_{oc}^{bur}$) suggests a mid-Proterozoic rate of ~0.5–2.5 Tmol C yr$^{-1}$, with a median of $1.10_{-0.64}^{+0.93}$ Tmol C yr$^{-1}$ (1$\sigma$) and a 95% credible interval of 0.20–3.41 Tmol C yr$^{-1}$ (Fig. 2a blue). These fluxes are well below estimates of marine organic carbon burial for the modern (10.5–13.3 Tmol C yr$^{-1}$) (Berner, 1982; Burdige, 2005; Muller-Karger et al., 2005), Holocene (11.4 Tmol C yr$^{-1}$) (Wallmann et al., 2012), and Quaternary (12.92 Tmol C yr$^{-1}$) (Wallmann et al., 2012). On the other hand, our model estimates the global burial rate of pyrite sulphur ($J_{py}^{bur}$) at $1.18_{-0.26}^{+0.26}$ Tmol S yr$^{-1}$ (1$\sigma$) with a 95% credible interval of 0.81–1.63 Tmol S yr$^{-1}$ (Fig. 2b blue), within a factor of 2 of the 'near-modern' (Quaternary average) ocean value of 1.2–1.6 Tmol S yr$^{-1}$ (Berner and Berner, 2012; Markovic et al., 2015).

Taken together, we estimate a combined rate of $O_2$ production of about 3.51 Tmol $O_2$ yr$^{-1}$, roughly 25% of the present value ($J_{O2}$ in Fig. 2c). We also find that the posterior distributions are largely unchanged for the 'High $O_2$' scenario (Fig. 2a,b red) in which a possible range of the atmospheric $O_2$ levels is set to 1–10% PAL (Table 1); in this case, the median value of $J_{oc}^{bur}$ and total $O_2$ production rate are 1.64$_{-0.98}^{+1.52}$ Tmol C yr$^{-1}$ and 4.06$_{-1.40}^{+1.91}$ Tmol yr$^{-1}$, respectively. In this study we designed the model to keep our arguments conservative. We therefore emphasize that our estimates derived above are likely an upper limit.

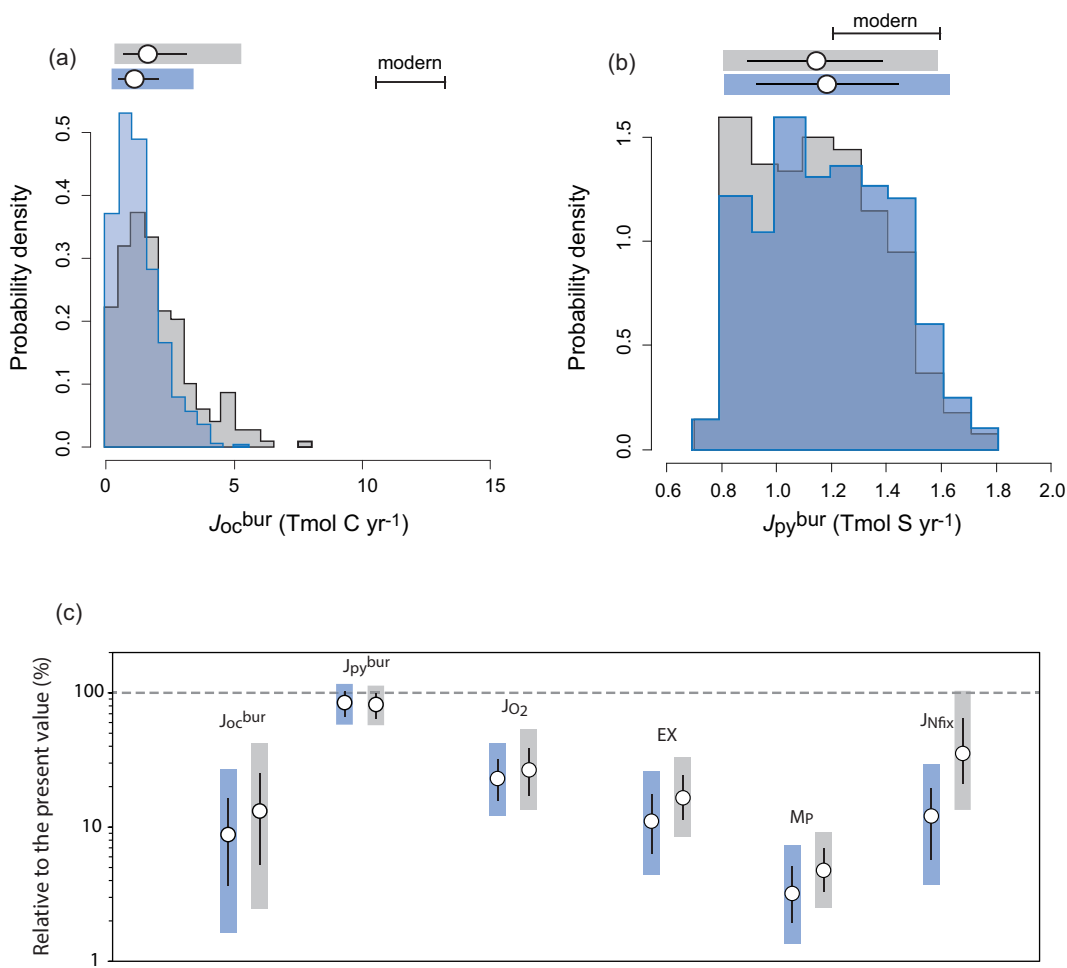

**FIGURE 2** Posterior probability distributions retrieved by our model. The burial rates of organic carbon ($J_{oc}^{bur}$: a) and pyrite sulphur ($J_{py}^{bur}$: b) from our 'Low $O_2$' (blue; $n = 621$) and 'High $O_2$' (red; $n = 308$) retrievals. Open circles represent median values. In both cases, error bar and grey shaded region denote 1$\sigma$ and 95% credible intervals, respectively. Double-headed arrows represent flux estimates for the modern or near modern oceans. (c) Model retrieval of mid-Proterozoic Earth system biogeochemistry. $J_{oc}^{bur}$ = burial rate of organic carbon. $J_{py}^{bur}$ = burial rate of pyrite sulphur. $J_{O2}$ = total rate of $O_2$ production. $M_P$ = marine phosphate inventory. $EX$ = export production of organic carbon. $J_{Nfix}$ = rate of nitrogen fixation. All values are normalized to the modern or near modern values.

Previous work has highlighted the issue of maintaining low $O_2$ fluxes in a pervasively reducing mid-Proterozoic ocean (Derry, 2015; Laakso and Schrag, 2014), given the enhanced preservation of organic matter in anoxic marine sediments (e.g., (Katsev and Crowe, 2015)). In our model, as in others (Laakso and Schrag, 2014; Reinhard et al., 2017), this can be attributed to strongly suppressed biological productivity ($EX$ in Fig. 2c). Given constraints on seawater $[SO_4^{2-}]$, our model requires MSR to be suppressed by decreasing the availability of organic matter for MSR, and this can be achieved by a suppression of biological productivity through a scarcity of the primary nutrient (phosphorus, P) in the ocean interior ($M_P$ in Fig. 2c)—our model indicates that the marine P inventory would have been <10% of the present oceanic level (POL), with a median estimate of $3.2_{-1.3}^{+1.9}$% POL ($1\sigma$) and a 95% credible interval of 1.4–7.3% POL for our 'Low $O_2$' scenario. Our estimate of the deep water phosphate concentration of ~0.1 $\mu$M (Fig. 3a) is within the range of 0.04–0.13 $\mu$M proposed by Jones et al. (Jones et al., 2015) from the P/Fe ratios of iron-rich chemical sediments deposited during the late Paleoproterozoic, but is based on a completely independent geologic constraint and implies that this condition extended well into the mid-Proterozoic.

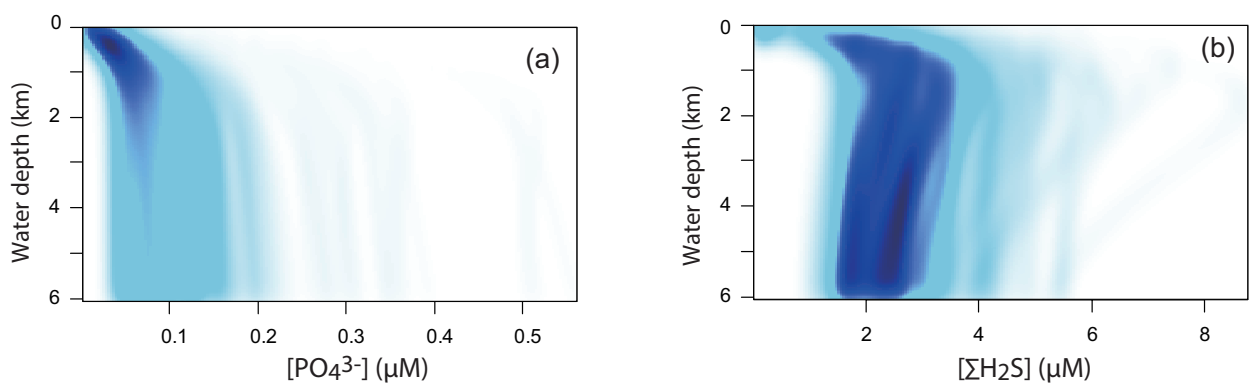

**FIGURE 3** Chemistry of the ocean interior on the mid-Proterozoic Earth. Probability distribution of phosphate (left; a) and total sulphide (right; b) in the low-mid latitude region for our 'Low $O_2$' retrieval. Frequency is shown in colour contours. Note that although some coastal marine settings would likely deviate above these concentrations, in general these are likely upper limits (see text).

However, we note that for our 'Low $O_2$' scenario biological productivity may also have been impacted by increased solar UV flux because total $O_3$ column depth decreases significantly below 1% PAL (Segura et al., 2003). Our model does not account for this possible harmful effect of UV on life. On the other hand, UV fluxes attenuate rapidly with depth in seawater, with typical 10% depths for DNA damage well below 10m in open ocean and coastal marine systems (Tedetti and Sempéré, 2006). In any case, our overarching result—strongly supressed marine new production— would only be reinforced by the impact of increased solar UV flux. We also note that our model estimates a broadly similar $M_P$ at 4.8$_{-1.4}^{+2.2}$% POL (1$\sigma$) for our 'High O$_2$' scenario, under which increased UV flux is not expected to play a significant role.

**3.2 The mid-Proterozoic Earth system**

Our model retrieval provides statistical constraints on a number of major Earth system processes during the mid-Proterozoic. First, we estimate that the global rate of nitrogen fixation in mid-Proterozoic oceans was substantially lower than the present value: our model predicts a globally integrated N fixation rate of 21$_{-11}^{+13}$ Tg N yr$^{-1}$ (1$\sigma$) with a 95% credible interval of 6.5–52 Tg N yr$^{-1}$. This value is a factor of ~7 below that estimated for the modern ocean (Eugster and Gruber, 2012; Groszkopf et al., 2012) ($J_{Nfix}$ in Fig. 2c), and arises straightforwardly from suppressed N loss via denitrification in the nitrate-lean and oligotrophic oceans such that very little biological nitrogen fixation is required to meet the nutrient N demands of the biosphere. The broader implication is that any long-term trace nutrient deficiency acting at a global scale on the mid-Proterozoic Earth (Anbar and Knoll, 2002; Reinhard et al., 2013; Scott et al., 2008) would need to have been very extreme to be the ultimate limiting factor on O$_2$ fluxes from the biosphere, consistent with a recently emerging view of the factors regulating Earth's oxygenic biosphere through time (Laakso and Schrag, 2018; Reinhard et al., 2017).

Second, limited organic matter production provides a simple explanation for the accumulated geochemical data indicating that euxinic waters have been a predominantly local phenomenon for most of Earth's history (Planavsky et al., 2011; Poulton and Canfield, 2011; Reinhard et al., 2013). We estimate relatively low total sulphide ($\Sigma H_2S$) concentrations in the ocean interior, with our retrieval yielding values on the order of ~1–3 $\mu$M for the abyssal oceans (Fig. 3b). It is important to bear in mind that this value is effectively a zonal/meridional average at each depth in our model such that deepwater H$_2$S concentrations well above this value are possible locally, particularly in coastal marine environments. On the other hand, the lack of a fully explicit Fe cycle in our model currently prevents us from diagnosing deep ocean [Fe$^{2+}$], and it is likely that for large regions of the ocean interior H$_2$S concentrations would be essentially negligible. Taken together, our results are consistent with the mid-Proterozoic global oceanic redox landscape inferred from geologic archives—globally ferruginous and locally euxinic (Planavsky et al., 2011; Poulton and Canfield, 2011)—and further suggest that this redox structure was controlled largely by limited flux of organic matter to the abyssal ocean (Reinhard et al., 2013; Scott et al., 2008). More granular constraints on the ocean redox landscape of the mid-Proterozoic Earth will require the development of more explicit Fe cycling in both low-order models like CANOPS and Earth system models of intermediate complexity (EMIC; (Claussen et al., 2002))

Third, our model retrieves relatively low atmospheric $CH_4$ mixing ratios due to suppression of methanogenesis and concomitant degassing of $CH_4$ from the ocean to the atmosphere (Fig. 4a,b). This can be attributed to the combined effect of limited organic matter availability for methanogenesis and aerobic and anaerobic $CH_4$ oxidation by $O_2$ and $SO_4^{2-}$ (AOM), respectively (Olson et al., 2016). We estimate median values of ~3.4 ppmv and ~13.2 ppmv for our 'Low $O_2$' and 'High $O_2$' scenarios, slightly higher than the present value by a factor of 2–7 and consistent with recent estimates from 3-D ocean biogeochemical models both with and without AOM (Daines and Lenton, 2016; Olson et al., 2016). Relatively low atmospheric $CH_4$ levels during the mid-Proterozoic reinforce recent suggestions (Olson et al., 2016) that $CH_4$ and $CO_2$ alone may not have been sufficient to buffer the mid-Proterozoic climate system, unless terrestrial and/or abiotic $CH_4$ sources were much greater than at present (Zhao et al., 2017). Moreover, low $CH_4$ mixing ratios in the atmosphere indicate a limited role of hydrogen escape to space in the mid-Proterozoic $O_2$ balance (see below).

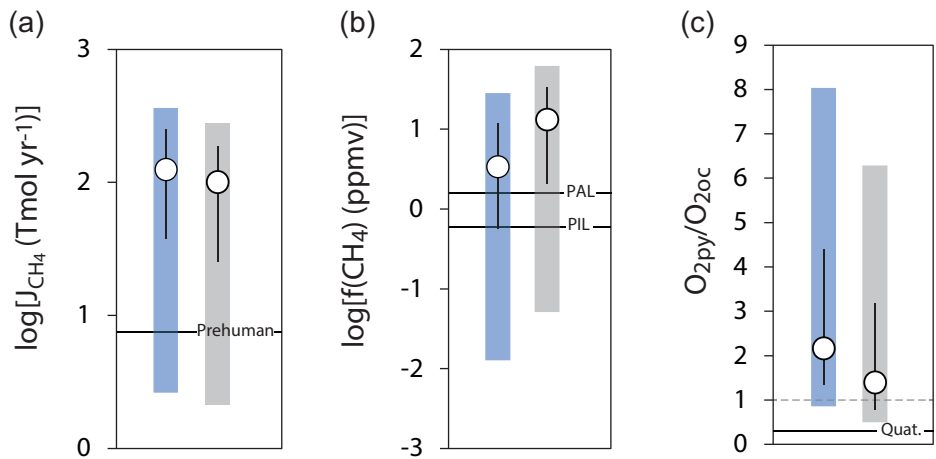

**FIGURE 4** Model retrieval of mid-Proterozoic Earth system biogeochemistry for our 'Low $O_2$' (blue) and 'High $O_2$' (red) scenarios. Open circles represent median values, while error bars and shaded regions represent $1\sigma$ and 95% credible intervals, respectively. (a) Degassing flux of $CH_4$ from the ocean to the atmosphere, the pre-human biospheric flux depicted as a horizontal solid line. (b) Atmospheric mixing ratio of $CH_4$. Preindustrial (PIL) and present (PAL) values are shown as horizontal solid lines. (c) Relative $O_2$ fluxes from pyrite sulphur burial ($O_{2py}$) and organic carbon burial ($O_{2oc}$). The Quaternary value depicted as a horizontal solid line.

Lastly, when we calculate the relative significance of pyrite burial and organic carbon burial in the $O_2$ production, our model shows that the rate of $O_2$ production via pyrite burial is comparable to, or larger than, organic carbon burial. We retrieve an $O_2$ production rate via pyrite burial that is ~1–8 times as large as the rate of $O_2$ production via organic carbon burial, with a median value of 2.2 for our 'Low $O_2$' scenario (Fig. 4c). Our 'High $O_2$' scenario also estimates a median value of 1.4 with a 95% credible interval of 0.5–6.3. These represent a remarkable contrast to Earth system biogeochemistry during most of Phanerozoic time, during which organic carbon burial has represented the dominant $O_2$ source (Berner and Raiswell, 1983). Our result is consistent with an independent argument based on the geological record of carbon and sulphur isotopes (Canfield, 2005), and suggests that the sulphur cycle would have been a critical component of the global $O_2$ budget during the mid-Proterozoic.

### 3.3 A global redox budget for the mid-Proterozoic

By embedding our estimates of $O_2$ production in a global redox balance framework (see Supplementary Methods), we can estimate the reductant fluxes from Earth's interior required to balance the global $O_2$ cycle across the range of parameters explored here (Table 2). We use a previously published parameterization for $O_2$ consumption associated with the oxidative weathering of organic matter (Daines et al., 2017; Lasaga and Ohmoto, 2002) (equation (S44)) and ferrous iron (Kanzaki and Murakami, 2016; Yokota et al., 2013) (equation (S45)) (see Supplementary Methods), combined with the $O_2$ production fluxes derived above, to estimate that an external reductant flux of 2.75 Tmol $O_2$ equivalents yr$^{-1}$ is required to satisfy global redox balance under the mid-Proterozoic Earth system state retrieved by our model for our 'Low $O_2$' scenario (Fig. 5a). This can be compared to a modern reductant flux to Earth's surface from volcanism, hydrothermal systems, and geologic/thermogenic $CH_4$ production of 1.5–2.1 Tmol $O_2$ equivalents yr$^{-1}$. Given that even the modern solid Earth reductant flux cannot be constrained with a precision better than ~1 Tmol $O_2$ equivalents yr$^{-1}$, it is clear that our 'Low $O_2$' retrieval is fully consistent with a closed, stable redox balance. This is also true for our 'High $O_2$' retrieval, in which 0.85 Tmol $O_2$ equivalents yr$^{-1}$ is sufficient to close the global redox budget (Table 2 and Fig. 5b).

## 3.4 Implications for the carbon isotope record

Geologic records of the carbon isotope composition of sedimentary carbonate rocks have often been used to estimate the relative fraction of carbon removed from the exogenic (ocean-atmosphere) system that is buried as organic carbon ($f_{org}$) (Broecker, 1970; Des Marais et al., 1992; Hayes and Waldbauer, 2006; Krissansen-Totton et al., 2015; Kump and Arthur, 1999; Schidlowski, 1988). Canonically, this approach has been used to suggest that the relative organic matter burial fraction has not changed significantly throughout Earth's history. In principle, this represents a challenge to models that invoke suppressed marine organic matter production and burial, given that overall rates of carbon flow through the exogenic system would have to change dramatically and in concert with changes in organic carbon productivity/burial in order to maintain a stable $f_{org}$. However, this analysis assumes that the geological carbon cycle is in steady state and that the carbon isotopic composition ($\delta^{13}C$) of the overall input to the exogenic system is identical to the canonical mantle value of -5‰, when in reality this parameter should scale with atmospheric $O_2$ as a result of $O_2$-dependent changes to the globally integrated rates of organic carbon weathering in the crust (e.g., (Bolton et al., 2006)).

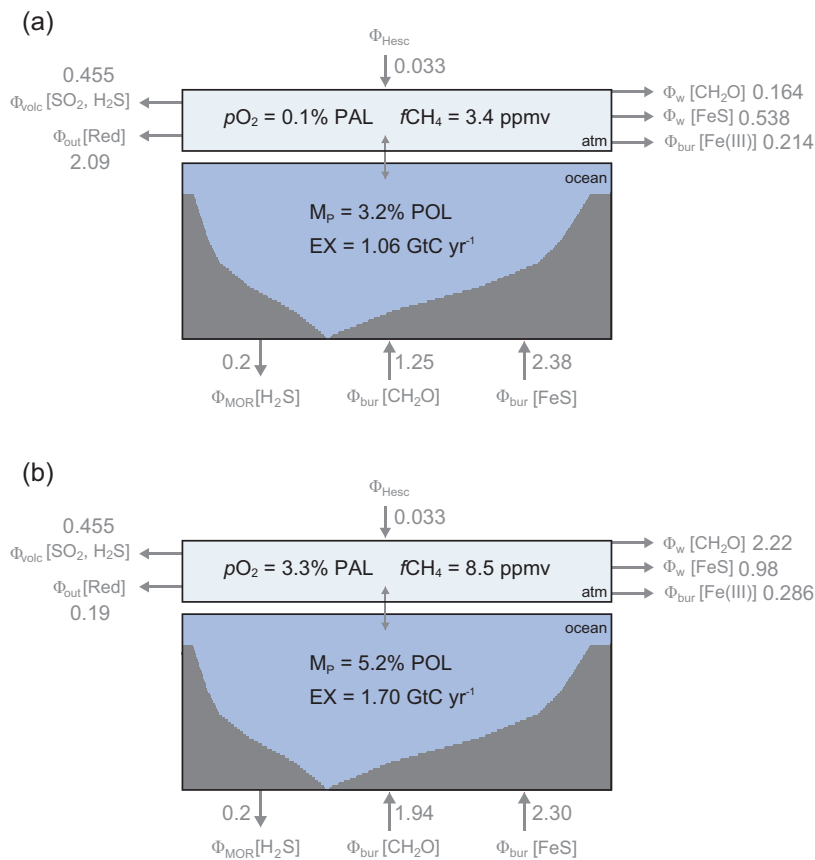

**FIGURE 5** Global redox budget analysis from our Monte Carlo analysis for our 'Low $O_2$' scenario (top: a) and 'High $O_2$' scenario (bottom: b) of Proterozoic Earth surface system. $\Phi$ represents the mean value of the flow of oxidizing power in the system in terms of Tmol $O_2$ equivalents yr$^{-1}$. $\Phi_{bur}(CH_2O)$ and $\Phi_{bur}(FeS)$ represent the input of oxidizing power via burial of organic matter and pyrite sulphur. $\Phi_{Hesc}$ denotes the flux of hydrogen escape to space, and 'w' is the oxidative weathering. $\Phi_{MOR}(H_2S)$ and $\Phi_{volc}(SO_2, H_2S)$ are the output of oxidizing power via input of reduced sulphur gases through volcanic activity, which are treated as a free parameter.

In contrast, our model predicts that oxidative weathering of organic matter would have been suppressed under mid-Proterozoic conditions, consistent with a number of recent models (Daines et al., 2017; Miyazaki et al., 2018). This results in $\delta^{13}C$ values for the integrated carbon flux to the ocean-atmosphere system that are significantly higher than the canonical mantle value (-0.3‰ and -1.1‰ for our 'Low $O_2$' and 'High $O_2$' scenarios, respectively; Table S8). When combined with rates of volcanic $CO_2$ outgassing from the mantle ($J_{mantle}$) and carbonate weathering ($J_{carb}^w$) that are 150% of the reference values ($J_{mantle}^* = 2$ Tmol C yr$^{-1}$ and $J_{carb}^* = 40$ Tmol C yr$^{-1}$ where * denotes the reference value), we estimate $f_{org}$ values of 1.2% and 4.3% for our 'Low $O_2$' and 'High $O_2$' scenarios, respectively (Table S8), compared to a canonical value of ~15–20% (e.g.,(Krissansen-Totton et al., 2015)). These results add to a growing body of work indicating that the quantitative scaling between the $\delta^{13}C$ values of marine sedimentary carbonate rocks and organic carbon burial fluxes from the exogenic system is somewhat obscure, particularly at low atmospheric $pO_2$ (Bjerrum and Canfield, 2004; Daines et al., 2017; Miyazaki et al., 2018; Schrag et al., 2013). Our model retrieval of the mid-Proterozoic Earth system is thus fully consistent with the existing carbon isotope record.

## 4 CONCLUSION

In summary, our statistical analysis of a novel biogeochemical model indicates that the rate of $O_2$ production in mid-Proterozoic oceans was lower than that in the modern ocean by as much as an order of magnitude or more, providing a compelling mechanism for explaining the protracted oxygenation of the Earth's atmosphere. We estimate the residence time of $O_2$ in the combined ocean-atmosphere system at ~10–300 kyr, implying either an intriguing but poorly understood series of rapid stabilizing feedbacks within the coupled carbon, nutrient, and oxygen cycles at Earth's surface or the potential for significant oscillation in ocean-atmosphere $O_2$ on relatively short timescales. In any case, our results provide a comprehensive and statistically robust picture of the mid-Proterozoic Earth system that is fully consistent with both current understanding of the coupled C-N-P-$O_2$-S cycles and existing constraints from the geologic record, and a baseline from which to examine how and why secular oxygenation of Earth's ocean-atmosphere system occurred during the late Neoproterozoic.

**SUPPORTING INFORMATION**

Additional Supporting Information may be found online in the supporting information tab for this article.